\documentclass{article} % For LaTeX2e
\usepackage[preprint]{colm2026_conference}

\usepackage{amsmath}
\usepackage{microtype}
\usepackage{hyperref}
\usepackage{url}
\usepackage{booktabs}
\usepackage{graphicx}
\usepackage{multirow}
\usepackage{xcolor}
\usepackage{tabularx}

\usepackage{lineno}

\definecolor{darkblue}{rgb}{0, 0, 0.5}
\hypersetup{colorlinks=true, citecolor=darkblue, linkcolor=darkblue, urlcolor=darkblue}

\newcommand{\taubench}{$\tau$-bench}
\newcommand{\realsim}{\textsc{RealUserSim}}
\newtheorem{definition}{Definition}

\title{RealUserSim: Bridging the Reality Gap in Agent Benchmarking via Grounded User Simulation}

\author{
\centerline{Ming Zhu\thanks{Correspondence to: \texttt{\{mingzhu, shelby.heinecke, huan.wang\}@salesforce.com}},~\textbf{Juntao Tan},~\textbf{Rithesh Murthy},~\textbf{Jielin Qiu},~\textbf{Liangwei Yang},}\\
\centerline{~\textbf{Wenting Zhao},~\textbf{Silvio Savarese},~\textbf{Shelby Heinecke},~\textbf{Huan Wang}}\\[2mm]
\centerline{Salesforce AI Research}
}

\begin{document}

\ifcolmsubmission
\linenumbers
\fi

\maketitle

\begin{abstract}
LLM-based user simulation is the primary mechanism for end-to-end agent evaluation, yet simulated users are poor proxies for real humans: unconstrained LLM defaults produce a \emph{Formalism Ceiling} (style match rates of 6--8\% against real users), while hand-crafted behavioral directives trigger \emph{Directive Amplification}, where models hyper-interpret instructions into unnatural behavioral extremes that vary dramatically across simulator models. We present \realsim{}, the first user simulation framework grounded in real behavioral data. From 14,000+ authentic human--LLM conversations (WildChat), we extract 7,275 executable behavioral profiles and use them to ground LLM simulators. A fidelity benchmark (PT3) on 600 conversations across 71+ domains with anti-leakage controls shows that grounded simulation raises match rate from 24.2\% to 45.3\% across five behavioral dimensions. Agent evaluation on \taubench{} with 6 simulator models and extensive analysis shows that grounded simulation acts as a realistic stress test, surfacing three failure mechanisms invisible to cooperative simulators (mean $-3.2\%$ to $-3.5\%$ task success degradation), while Directive Amplification in existing benchmarks produces unrealistic behavior that compromises the validity of agent evaluation.
\end{abstract}

%============================================================================
\section{Introduction}
\label{sec:intro}

Large Language Models (LLMs) are increasingly deployed as autonomous, user-facing agents in task-oriented domains such as customer service, technical support, and e-commerce~\citep{yao2024tau, barres2025tau2}. Accurately benchmarking these agents is critical, yet live user studies are expensive and difficult to scale. The community has therefore turned to LLM-based user simulation as the primary mechanism for end-to-end agent evaluation~\citep{yao2024tau, ren2024bases, bougie2025simuser}. However, growing evidence suggests that simulated users are poor proxies for real humans: they are excessively cooperative, stylistically uniform, and lack realistic frustration or ambiguity~\citep{seshadri2026lost, zhou2026sim2real}, inflating agent performance and obscuring deployment-critical failure modes.

We identify two root causes of this \textbf{Reality Gap}. First, unconstrained LLM simulators default to a ``cooperative assistant'' persona: grammatically correct, well-punctuated, verbose text regardless of the target user's actual style. We show that without behavioral grounding, style-related match rates against real users are near-floor (6--8\%), a phenomenon we term the \textbf{Formalism Ceiling}. Second, hand-crafted behavioral directives (e.g., ``you are an angry customer'') trigger \textbf{Directive Amplification}: models hyper-interpret instructions into unnatural behavioral extremes, complete with stage directions (e.g., \texttt{*sniffling*}) and hyper-structured multi-request patterns that no real human would produce. We show that the same directive produces dramatically different behavior across models, making benchmark scores not comparable across simulator models when directives are present.

Concurrent work has diagnosed aspects of this problem. \citet{seshadri2026lost} conduct human studies showing that LLM-simulated users produce different conversational patterns and failure modes than real humans, with success rates fluctuating by up to 9 percentage points across simulator models. \citet{zhou2026sim2real} evaluate 31 LLM simulators against 451 human participants and find that higher model capability does not yield more faithful simulation. Both works conclude that human validation is essential. However, these studies are primarily diagnostic: they characterize the sim-to-real gap but do not provide a constructive framework for closing it. In contrast, \realsim{} offers a prescriptive solution: we extract behavioral profiles from real user data and use them to ground LLM simulators, demonstrating measurable fidelity improvements across multiple behavioral dimensions. While prior persona-grounded approaches rely on synthetic or fictional personas~\citep{wang2024rolellm, shao2023characterllm, park2023generative, wang2025deeppersona}, \realsim{} is the first framework to ground simulation in real user behavioral data extracted from authentic human--LLM interactions.

We present \realsim{}, a grounded user simulation framework comprising two core modules: (1)~a profile construction pipeline that extracts behavioral models from real user data, and (2)~a grounded simulation pipeline with fidelity evaluation via the Paired Trajectory Turing Test (PT3). Using 14,000+ authentic conversations from WildChat~\citep{zhao2024wildchat}, we construct 7,000+ executable behavioral profiles containing demographic summaries and linguistic style definitions grounded in observed data. Our contributions are:

\begin{itemize}
    \item We present the first user simulation framework grounded in real behavioral data, extracting 7,275 executable profiles from authentic human--LLM interactions (WildChat) to replace synthetic or fictional personas in agent evaluation.

    \item We introduce a user simulation fidelity benchmark (PT3) that measures faithfulness across five behavioral dimensions. On 600 conversations spanning 71+ domains with anti-leakage controls, grounded user simulation raises match rate from 24.2\% to 45.3\%, breaking through the Formalism Ceiling where ungrounded simulators score 6--8\% on style dimensions.

    \item Through agent evaluation on \taubench{} with 6 simulator models and extensive analysis, we show that grounded user simulation acts as a realistic stress test that surfaces three failure mechanisms invisible to cooperative simulators (mean $-3.2\%$ to $-3.5\%$ degradation), bridging the reality gap in agent benchmarking. We further show that Directive Amplification in existing benchmarks produces unrealistic simulator behavior that compromises the validity of agent evaluation.
\end{itemize}

%============================================================================
\section{Related work}
\label{sec:related}

\paragraph{Persona-based user simulation.}
User simulation has evolved from rule-based~\citep{schatzmann2006survey} and neural~\citep{kreyssig2018neural} approaches to LLM-driven role-playing agents and generative simulacra~\citep{shanahan2023roleplay, chen2024persona, park2023generative}. Recent work develops persona simulation architectures through character-level fine-tuning and benchmarking~\citep{wang2024rolellm, shao2023characterllm}, persona consistency via reinforcement learning and structured reasoning~\citep{abdulhai2025consistently, du2026her, kim2026picon}, and scalable persona generation~\citep{wang2025deeppersona, wang2025personaevolve}. Evaluation frameworks assess persona fidelity through dynamic metrics~\citep{samuel2025personagym} and knowledge-grounded simulation~\citep{shea2025sage}. However, prior persona-grounded dialogue focuses on knowledge-based personas~\citep{zhang2018personalizing} or utterance-level authenticity in task-oriented settings~\citep{wang2025usp}. \realsim{} instead grounds simulation in \emph{real user behavioral data} from authentic human--LLM interactions~\citep{zhao2024wildchat}, focusing on \emph{how} users communicate rather than \emph{what} they know, with a fidelity benchmark measuring faithfulness across multiple behavioral dimensions.

\paragraph{Interactive agent benchmarking.}
\taubench{}~\citep{yao2024tau} established multi-turn user--agent evaluation for tool-augmented agents, extended by $\tau^2$-bench~\citep{barres2025tau2} to dual-control settings. Benchmarks now span collaborative reasoning~\citep{sun2025collab, zhou2025sweetrl}, data analysis and search~\citep{li2025ida, deng2025interactcomp}, memory-driven conversations~\citep{bian2026realmem, jiayang2026amemgym}, adversarial robustness~\citep{jiang2026agentlab}, and domain-specific evaluation including web search, recommendation, product search, and policy adherence~\citep{ren2024bases, bougie2025simuser, ye2024proclare, shang2025agentrecbench, nakash2025redteam}. However, these benchmarks either use unconstrained LLM defaults or hand-crafted behavioral directives as user simulators, both of which we show produce unrealistic interaction patterns. \realsim{} addresses this gap by grounding the user simulator in real behavioral data, surfacing agent brittleness to authentic communication style variation.

%============================================================================
\section{Methodology}
\label{sec:method}

The \realsim{} framework consists of two core modules: (1)~constructing behavioral profiles from real user data, and (2)~grounded user simulation with fidelity evaluation.

\subsection{Profile construction from real users}
\label{sec:profiles}

We construct \textbf{7,275 Conversational Profiles} from WildChat-4.8M~\citep{zhao2024wildchat}, a large-scale dataset of real human--LLM interactions. Starting from 3.2M conversations, we filter to 21,637 multi-turn ($\geq$3 turns) English GPT-4o trajectories across 7,311 unique users (see Appendix~\ref{app:pipeline} for the full pipeline). Each profile contains a \emph{linguistic style profile} and a \emph{demographic profile}.

\paragraph{User persona profiling.}
\label{sec:linguistic}
For each user, GPT-4o analyzes their conversation history to produce an \textbf{Executable Persona Manual}: up to 15 ``Command + Example'' pairs capturing recurring communication patterns. Each command is a direct instruction for an LLM simulator; each example is quoted from the user's actual messages:

\begin{quote}
\small
\texttt{Command: Frequently repeat words or phrases for emphasis.}\\
\texttt{Examples: "the most the most famous food";}\\\texttt{\phantom{Examples: }"there are there are huge differences"}\\[4pt]
\texttt{Command: Use "hmm" or "ahh" to indicate thinking or hesitation.}\\
\texttt{Examples: "hmm i think the older people"; "hmmm if we eat veegtables"}\\[4pt]
\texttt{Command: Request additional examples or options when unsatisfied.}\\
\texttt{Examples: "share a more short caption."; "share 10 more"}
\end{quote}

This format is critical: rather than abstract descriptions (e.g., ``informal writer''), executable instructions constrain the simulator to reproduce specific, observed behaviors. The commands describe \emph{what} pattern to reproduce; the examples calibrate \emph{to what degree}. We generate 7,275 persona profiles (99.5\% of users).

\paragraph{Demographic profiling.}
\label{sec:demo_profiling}
Each profile is augmented with demographic attributes (age, education, gender, occupation, etc.) via two phases: (1)~GPT-4o extracts explicit self-disclosures from conversation text, aggregated via majority voting across trajectories; (2)~a hybrid inference pipeline predicts missing fields from conversational cues (see Appendix~\ref{app:pipeline} for validation). The three sources (persona profiles, extracted demographics, inferred demographics) are merged into 7,275 unified profiles.

\begin{figure*}[t]
\centering
\includegraphics[width=\textwidth]{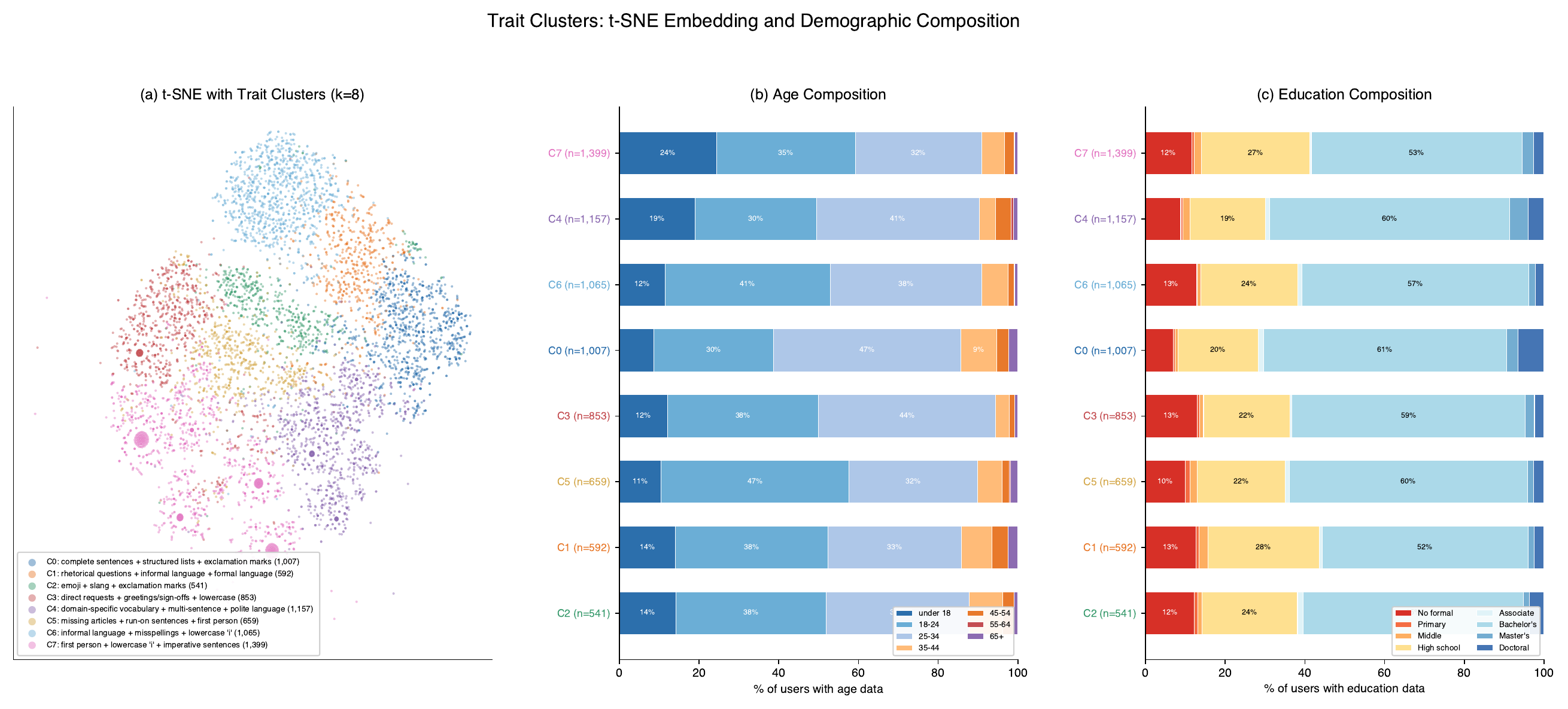}
\caption{Linguistic trait clusters across 7,273 users. (a)~t-SNE projection of 48-dimensional binary trait vectors with $k$-means clustering ($k{=}8$). Each cluster represents a behavioral archetype (e.g., C7: Everyday Casual, C4: Formal Expert, C6: Informal Texter, C0: Organized Writer). (b,~c)~Age and education composition per cluster, showing systematic demographic shifts: casual/informal clusters skew younger and lower-education, while formal/structured clusters skew older and higher-education.}
\label{fig:tsne_clusters}
\end{figure*}

\paragraph{Profile diversity.}
Figure~\ref{fig:tsne_clusters} shows a t-SNE projection of all 7,273 users based on 48 binary linguistic traits, revealing 8 distinct behavioral archetypes---from \emph{Everyday Casual} (terse, lowercase, imperative) to \emph{Formal Expert} (domain vocabulary, polite, multi-sentence). The demographic composition of these clusters differs systematically: younger and lower-education users concentrate in casual and informal clusters, while older, higher-education users concentrate in formal and structured clusters. This confirms that the extracted profiles capture genuine, structured behavioral diversity rather than noise.

\subsection{Grounded user simulation and fidelity evaluation}
\label{sec:simulation}

We evaluate profile-grounded user simulation by synthesizing parallel conversations and measuring behavioral fidelity against real ones. For each real conversation, GPT-4o first extracts a \emph{problem description} and \emph{solution conditions} from the first 10 messages as a shared task specification. A simulated user then interacts with a GPT-4o agent to achieve this goal: in \textbf{Baseline} mode, the simulator receives only the goal; in \textbf{With Profile} mode, it additionally receives the user's demographic summary and linguistic profile. The system prompt includes a critical anti-normalization constraint: ``Do not `clean up' the writing. If the commands require poor grammar and frequent typos, your response must be equally messy.'' Conversations are capped at 9 agent messages. Finally, an LLM judge (GPT-4o, temperature $\approx$ 0) compares the synthetic and original conversations using the Paired Trajectory Turing Test (PT3, described below).

\subsubsection{The Paired Trajectory Turing Test (PT3)}
\label{sec:fidelity}

\begin{definition}[Paired Audit]
Let $\mathcal{T}_h$ be a historical trajectory between a human user $H$ and an agent $A$. Let $\mathcal{T}_s$ be a synthetic trajectory generated by a simulated user grounded in the profile of $H$ interacting with the same agent $A$ (or a high-fidelity proxy) under an identical scenario.
\end{definition}

The judge is presented with $\mathcal{T}_h$ and $\mathcal{T}_s$ without labels, and evaluates consistency across five dimensions of conversational identity (Table~\ref{tab:fidelity_dims}), assigning a binary match/no-match verdict for each. Termination signals are stripped to avoid evaluation artifacts. The overall Fidelity Index is the mean match rate across all dimensions (defaulting to equal weights).

\begin{table*}[t]
\centering
\small
\caption{Five dimensions of conversational identity used in the PT3 fidelity evaluation.}
\label{tab:fidelity_dims}
\begin{tabularx}{\textwidth}{@{}lX@{}}
\toprule
\textbf{Dimension} & \textbf{Description} \\
\midrule
Persona \& Affective Traits & Demeanor, emotional state, patience level, personality cues \\
Linguistic Style \& Mechanics & Vocabulary, phrasing, formality, typos, message length patterns \\
Tech Competency \& Knowledge & Domain expertise, terminology usage, depth of questions \\
Interaction \& Data Flow & Information-sharing habits (scattered vs.\ dense), questioning style \\
Pacing \& Action Sequencing & Turn length, topic transitions, conversation conclusion patterns \\
\bottomrule
\end{tabularx}
\end{table*}

\subsubsection{Benchmark design}
\label{sec:benchmark_design}

The benchmark comprises \textbf{6 evaluation subsets} totaling 600 conversations from 504 unique users: five domain-specific subsets (Technology \& IT, Business \& Finance, Medical \& Health, E-commerce, Travel \& Hospitality; 100 conversations each) and one mixed-domain subset spanning 71 domains (100 conversations). Each subset enforces a maximum of 2 conversations per user to ensure diversity.

\paragraph{Anti-leakage cleaning.}
Since profiles are consolidated from a user's full conversation history, examples may originate from the test conversation itself. We apply per-test-case cleaning: (1)~single-conversation users have all examples stripped, retaining commands only; (2)~multi-conversation users have examples tagged with the test conversation removed; (3)~users whose profile was built entirely from other conversations require no cleaning. Across the benchmark, 150 cases are stripped entirely, 380 have specific examples removed, and 70 require no change.

%============================================================================
\section{Experiments}
\label{sec:experiments}

We evaluate \realsim{} in two settings: (1) a user simulation fidelity benchmark on WildChat conversations, and (2) an end-to-end agent evaluation on \taubench{}.

\subsection{User simulation fidelity evaluation}
\label{sec:fidelity_setup}

We construct 6 evaluation subsets of 100 conversations each (600 total) from WildChat users with consolidated profiles, covering 504 unique users across 71+ domains. Five subsets are domain-specific (Technology \& IT, Medical \& Health, Business \& Finance, E-commerce, Travel \& Hospitality); the sixth is mixed-domain, spanning 71 domains. Conversations are quality-filtered to ensure genuine multi-turn dialogue ($\geq$3 user turns, quality score $\geq$10/15). Profiles undergo per-test-case anti-leakage cleaning (Section~\ref{sec:benchmark_design}). Full construction criteria and statistics are in Appendix~\ref{app:fidelity_setup}.

We compare two conditions: (1)~\textbf{Baseline} (task specification only) and (2)~\textbf{With Profile} (task specification + cleaned demographic summary and linguistic profile), yielding 3,000 PT3 judgments per condition. All models use GPT-4o; the simulated user and agent use temperature 0.7, and the judge uses temperature $\approx$ 0.

\subsection{End-to-end agent evaluation}
\label{sec:agent_setup}

We evaluate on \taubench{}~\citep{yao2024tau}, a customer-service agent evaluation framework with Airline (50 tasks) and Retail (114 tasks) domains. All runs use GPT-4.1 as the agent (temperature 0.0). We evaluate 6 user simulator models: GPT-4o, GPT-5-mini, GPT-5, Llama~3-70b, gpt-oss-20b, and Claude~3 Sonnet.

We compare three conditions: (1)~\textbf{Original:} unmodified \taubench{} with embedded behavioral directives; (2)~\textbf{No Persona (NP):} task-only scenarios with directives stripped via LLM-based separation (Appendix~\ref{app:separation}); and (3)~\textbf{Real Persona:} injection of grounded WildChat profiles, randomly sampled per task, averaged across 3 independent runs with different random assignments. Both NP and Real Persona use identical task-only instructions, ensuring they differ only in whether a persona block is present. Additional analyses of demographic pools and a synthetic Perfect User are in Appendix~\ref{app:additional_agent}.

%============================================================================
\section{Results and analysis}
\label{sec:results}

\subsection{User simulation fidelity}
\label{sec:fidelity_results}

We evaluate simulation fidelity using PT3 (Section~\ref{sec:fidelity}) on 600 WildChat conversations across 6 evaluation subsets, comparing Baseline (no profile) and With Profile conditions against real human trajectories. All profiles are cleaned to prevent data leakage (Section~\ref{sec:benchmark_design}).

\paragraph{Profiles consistently improve behavioral fidelity.}
Table~\ref{tab:fidelity_results} presents dimension-level aggregate results. The overall match rate increases from 24.2\% to 45.3\% (+21.1 points) across all 600 conversations. The improvement is consistent across all four style-sensitive dimensions: Persona \& Affective Traits (+31.8), Interaction \& Data Flow (+28.3), Pacing \& Action Sequencing (+25.5), and Linguistic Style \& Mechanics (+20.0). Tech Competency \& Knowledge is near-ceiling at baseline (93.3\%) and shows no change with profiles ($-$0.2), confirming that profiles model \emph{communication style}, not domain knowledge.

\begin{table}[t]
\centering
\small
\caption{PT3 fidelity evaluation across 600 WildChat conversations. Match rate (\%) indicates the proportion of cases where the LLM judge finds the simulated user indistinguishable from the real user. Top: aggregate by dimension. Bottom: by evaluation subset ($n{=}100$ each).}
\label{tab:fidelity_results}
\begin{tabular}{@{}lccc@{}}
\toprule
& \textbf{Baseline} & \textbf{With Profile} & $\boldsymbol{\Delta}$ \\
\midrule
\multicolumn{4}{@{}l}{\textit{By dimension}} \\
\quad Persona \& Affective Traits       & 7.2  & 39.0 & +31.8 \\
\quad Linguistic Style \& Mechanics      & 6.2  & 26.2 & +20.0 \\
\quad Tech Competency \& Knowledge       & 93.3 & 93.2 & $-$0.2  \\
\quad Interaction \& Data Flow           & 7.7  & 36.0 & +28.3 \\
\quad Pacing \& Action Sequencing        & 6.5  & 32.0 & +25.5 \\
\midrule
\multicolumn{4}{@{}l}{\textit{By subset}} \\
\quad Technology \& IT         & 21.4 & 38.0 & +16.6 \\
\quad Business \& Finance      & 22.6 & 47.4 & +24.8 \\
\quad Medical \& Health        & 19.2 & 39.4 & +20.2 \\
\quad E-commerce               & 40.4 & 59.4 & +19.0 \\
\quad Travel \& Hospitality    & 17.8 & 40.6 & +22.8 \\
\quad Mixed-Domain (71 domains) & 23.6 & 46.8 & +23.2 \\
\midrule
\textbf{Overall}        & \textbf{24.2} & \textbf{45.3} & \textbf{+21.1} \\
\bottomrule
\end{tabular}
\end{table}

\paragraph{Gains are robust across domains.}
Table~\ref{tab:fidelity_results} (bottom) shows per-subset results. Profile injection produces gains in all 6 subsets, with deltas ranging from +16.6 (Technology \& IT) to +24.8 (Business \& Finance). The mixed-domain subset (+23.2), which spans 71 domains, produces a delta comparable to the domain-specific average (+20.7), confirming that profile utility generalizes across conversational contexts rather than depending on within-domain homogeneity.

\paragraph{The Formalism Ceiling.}
Without a profile, match rates for style-related dimensions are near-floor (6--8\%), revealing a severe ``Formalism Ceiling'': the LLM defaults to grammatically correct, well-punctuated, verbose text regardless of the target user's actual style. The Executable Persona Manual format (Section~\ref{sec:linguistic}) is critical to breaking through this ceiling.

\paragraph{Linguistic Style \& Mechanics remains the hardest dimension.}
Despite profiles, Linguistic Style \& Mechanics achieves the lowest with-profile match rate (26.2\%). GPT-4o struggles to produce genuinely messy or idiosyncratic text even with explicit instructions. E-commerce is an exception (45.0\%), likely because e-commerce users have formulaic communication patterns (product descriptions, SEO-oriented phrasing) that are easier to reproduce. The proportion of single-conversation users, who lose all examples during anti-leakage cleaning, directly predicts Linguistic Style \& Mechanics performance: Travel \& Hospitality (41\% single-conversation users) achieves only 17.0\%, while Business \& Finance (24\%) achieves 30.0\%.

\paragraph{Tech Competency \& Knowledge occasionally regresses in specialized domains.}
In Technology \& IT, Tech Competency \& Knowledge drops from 99.0\% to 89.0\% ($-$10.0). Consolidated profiles represent a user's \emph{average} behavior across conversations; commands like ``Use medical and technical terminology accurately'' cause the simulator to overshoot the expertise level of the specific test conversation. In the mixed-domain subset, Tech Competency \& Knowledge shows a mild gain (+5.0), suggesting that across diverse domains, profiles provide positive calibration rather than overshoot.

\paragraph{Qualitative examples.}
Table~\ref{tab:qualitative} shows contrastive examples across three user types. In each case, the Baseline transforms the user's distinctive style into polished formal prose, while profile injection preserves characteristic patterns: abbreviations and missing punctuation (informal student), oral disfluencies and filler words (ESL speaker), and idiosyncratic formatting such as period-prefixed questions (medical student). The Command/Examples format serves a critical calibration function: commands alone can cause overshooting into caricature, while the accompanying examples anchor the degree of each trait to the user's actual behavior.

\begin{table*}[t]
\centering
\small
\caption{Contrastive examples of user simulation fidelity. The Baseline normalizes diverse communication styles into polished prose; profile injection preserves distinctive patterns.}
\label{tab:qualitative}
\begin{tabularx}{\textwidth}{@{}l>{\raggedright\arraybackslash}X>{\raggedright\arraybackslash}X>{\raggedright\arraybackslash}X@{}}
\toprule
\textbf{User Type} & \textbf{Original} & \textbf{Baseline} & \textbf{With Profile} \\
\midrule
Informal student &
\emph{``Class 6th ncert civics questions can u list three things that the government does which have not been mentioned''} &
\emph{``Hi there! I'm looking to understand more about the roles of government\ldots Could you help me identify three such functions?''} &
\emph{``hey can u tell me 3 things the government does that arent in the class 6 civics book? easy words pls''} \\
\midrule
ESL speaker &
\emph{``well i believe there are many differences between between two types of study for example for example hmm through face to face teaching, hmm students can students have\ldots''} &
\emph{``Hi there! I have a piece of text that I'd like to revise for clarity and accuracy\ldots''} &
\emph{``well hmm i need a lil help with something can you help me revise a context i got hmm it needs to be more clear and accurate you know\ldots''} \\
\midrule
Medical student &
\emph{``.which of the following therapies is most likely to prolong survival in COPD? a) atenolol b) enalapril c) Oxygen d) prednisone e) theophylline''} &
\emph{``Hi there! I recently asked a question about `Pulsus bisferiens'\ldots I'm looking for a similar type of question\ldots''} &
\emph{``.the pulse characteristic known as `water-hammer pulse' is associated with which of the following conditions: a) mitral stenosis b) aortic regurgitation\ldots''} \\
\bottomrule
\end{tabularx}
\end{table*}

\subsection{Agent evaluation}
\label{sec:agent_eval}

\subsubsection{Main results}
\label{sec:main_results}

\begin{figure*}[t]
\centering
\includegraphics[width=\textwidth]{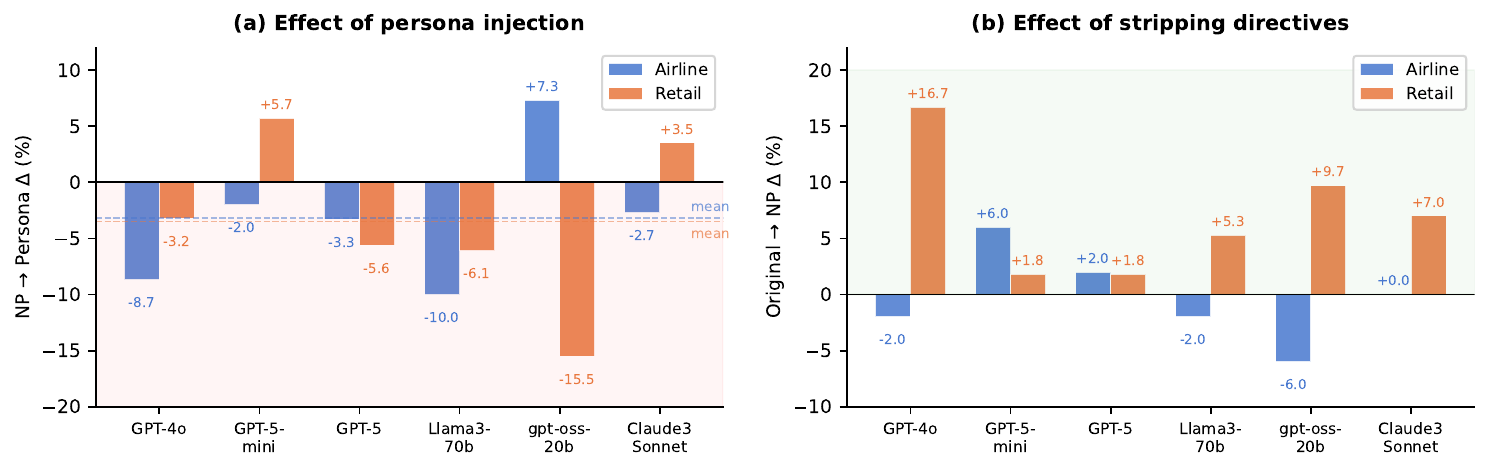}
\caption{Two key effects on agent task success rate across 6 simulator models. (a)~Persona injection (NP$\to$Persona) degrades most models in both domains (mean $-3.2\%$ Airline, $-3.5\%$ Retail), confirming that real user communication styles add meaningful friction. (b)~Stripping behavioral directives (Original$\to$NP) consistently improves Retail performance (all 6 models, up to $+16.7\%$), indicating that hand-crafted directives introduce artificial difficulty.}
\label{fig:persona_effect}
\end{figure*}

Table~\ref{tab:main_results} presents agent task success rates across all conditions, with persona results averaged over 3 independent random seeds. Figure~\ref{fig:persona_effect} visualizes the two key effects.

\begin{table*}[t]
\centering
\small
\caption{Agent task success rates (\%) across Airline ($n=50$) and Retail ($n=114$) domains. Persona averages 3 independent runs with different random persona--task assignments. $\Delta$ shows change from NP baseline.}
\label{tab:main_results}
\begin{tabularx}{\textwidth}{@{}l*{4}{>{\centering\arraybackslash}X}*{4}{>{\centering\arraybackslash}X}@{}}
\toprule
 & \multicolumn{4}{c}{\textbf{Airline} ($n=50$)} & \multicolumn{4}{c}{\textbf{Retail} ($n=114$)} \\
\cmidrule(lr){2-5} \cmidrule(lr){6-9}
\textbf{Model} & Orig & NP & Persona & $\Delta$ & Orig & NP & Persona & $\Delta$ \\
\midrule
GPT-4o         & 52.0 & 50.0 & 41.3 & $-$8.7  & 60.5 & 77.2 & 74.0 & $-$3.2 \\
GPT-5-mini     & 42.0 & 48.0 & 46.0 & $-$2.0  & 49.1 & 50.9 & 56.6 & +5.7 \\
GPT-5          & 46.0 & 48.0 & 44.7 & $-$3.3  & 75.4 & 77.2 & 71.6 & $-$5.6 \\
Llama~3-70b    & 56.0 & 54.0 & 44.0 & $-$10.0 & 64.0 & 69.3 & 63.2 & $-$6.1 \\
gpt-oss-20b    & 44.0 & 38.0 & 45.3 & +7.3    & 53.5 & 63.2 & 47.7 & $-$15.5 \\
Claude~3 Sonnet & 52.0 & 52.0 & 49.3 & $-$2.7 & 65.8 & 72.8 & 76.3 & +3.5 \\
\midrule
\textbf{Mean}  & 48.7 & 48.3 & 45.1 & $-$3.2  & 61.4 & 68.4 & 64.9 & $-$3.5 \\
\bottomrule
\end{tabularx}
\end{table*}

\paragraph{Persona injection degrades most models in both domains.} The mean NP$\to$Persona delta is $-3.2\%$ in Airline (5/6 models degrade) and $-3.5\%$ in Retail (4/6 models degrade), confirming that authentic human communication patterns add meaningful difficulty. The degradation is strongly model-dependent: Llama~3-70b shows the largest consistent drop ($-10.0\%$ Airline, $-6.1\%$ Retail), while gpt-oss-20b shows extreme domain dependence ($+7.3\%$ Airline, $-15.5\%$ Retail). Only GPT-5-mini and Claude~3 Sonnet benefit in Retail ($+5.7\%$ and $+3.5\%$).

\paragraph{Stripping directives helps Retail dramatically.} Removing behavioral directives (Original$\to$NP) consistently improves Retail performance across all 6 models (up to $+16.7\%$ for GPT-4o), indicating that hand-crafted directives introduce artificial difficulty. In Airline, the effect is near-zero (mean $-0.3\%$). The asymmetry arises because Retail tasks involve complex multi-step operations where adversarial directives compound operational complexity, while Airline directives mix helpful task-flow cues with harmful adversarial pressure that roughly cancel out.

\subsubsection{Failure mechanisms}
\label{sec:failure}

We examined trajectories where NP succeeded but Persona failed, identifying three failure mechanisms (Table~\ref{tab:failure_mechanisms}). The degradation follows a consistent pattern: personas compete with scenario instructions for control of user behavior, and style often wins over task logic; real user styles are optimized for natural conversation, not benchmark tasks requiring precise information transfer; and the agent is a co-participant, as many errors involve the agent misinterpreting the persona's style. Not all effects are negative---terse personas occasionally produce more unambiguous instructions that improve task completion.

\begin{table*}[t]
\centering
\small
\caption{Three failure mechanisms from persona injection, identified by comparing NP (pass) vs.\ Persona (fail) trajectories.}
\label{tab:failure_mechanisms}
\begin{tabularx}{\textwidth}{@{}>{\raggedright\arraybackslash}p{2cm}>{\raggedright\arraybackslash}p{3.8cm}X@{}}
\toprule
\textbf{Mechanism} & \textbf{Description} & \textbf{Example} \\
\midrule
Information loss & Terse persona omits critical task details & Airline Task~42: NP says \emph{``cancel FDZ0T5 and HSR97W \ldots don't modify other passengers.''} Terse persona says \emph{``proceed with cancelling the reservations as listed.''} Agent over-cancels 4 instead of 2. \\
\addlinespace
Agent misinterpretation & Aggressive style triggers wrong agent action & Airline Task~27: NP requests compensation, receives \$150 certificate. ALL-CAPS persona writes \emph{``GIMME A REFUND PLS''}---agent interprets as cancellation request. \\
\addlinespace
Compliance failure & Agreeable persona fails adversarial scenario & Airline Task~47: Must insist on refund 5$\times$ then give up \emph{without cancelling}. Hesitant persona says \emph{``ahh yes please go ahead and cancel it''} on first attempt. \\
\bottomrule
\end{tabularx}
\end{table*}

\subsubsection{Directive Amplification and model behavior}
\label{sec:amplification}

\paragraph{The same directive produces different behavior across models.} The directive ``You are extremely distraught'' causes Llama~3-70b to produce stage directions (\emph{``*sobbing* Oh, hi\ldots{} *choking back tears*''}), while GPT-4o simply writes: \emph{``I'm really upset right now.''} Table~\ref{tab:directive_sensitivity} quantifies this: models differ dramatically in sensitivity, from GPT-5-mini (message length doubles, frustration spikes) to GPT-4o (near-immune). This means benchmark difficulty changes depending on which simulator model is used, making scores \emph{not comparable across simulator models} when behavioral directives are present.

\begin{table*}[t]
\centering
\small
\caption{Behavioral metrics with and without directives. Roleplay markers are asterisk-bracketed actions (e.g., \texttt{*sobbing*}). Theatrical behavior drops sharply when directives are stripped.}
\label{tab:directive_sensitivity}
\begin{tabular}{@{}llcc@{}}
\toprule
\textbf{Model} & \textbf{Metric} & \textbf{With Directives} & \textbf{Without (NP)} \\
\midrule
\multirow{2}{*}{Llama~3-70b} & Roleplay markers & 8.6\% & \textbf{0.9\%} \\
 & Stage directions & 1.5\% & \textbf{0.0\%} \\
\midrule
\multirow{2}{*}{GPT-5-mini} & Frustration words & 16.7\% & \textbf{2.1\%} \\
 & Avg message length & 1218 chars & \textbf{514 chars} \\
\midrule
\multirow{2}{*}{gpt-oss-20b} & Roleplay markers & 19.4\% & \textbf{10.9\%} \\
 & Frustration words & 8.2\% & \textbf{2.0\%} \\
\midrule
\multirow{2}{*}{GPT-4o} & Roleplay markers & 0.0\% & 0.0\% \\
 & Frustration words & 1.2\% & 0.9\% \\
\bottomrule
\end{tabular}
\end{table*}

\paragraph{Model-specific behaviors.} Claude~3 Sonnet is the only model that consistently improves with persona injection in Retail ($+3.5\%$): its default behavior is too accommodating, proactively revealing information it should withhold, while a terse persona overrides this and correctly holds firm. Conversely, gpt-oss-20b shows the most extreme Retail degradation ($-15.5\%$) because it ignores system messages entirely, making persona instructions ineffective. Without directives (NP condition), models cluster into concise/conversational (GPT-4o, Claude~3 Sonnet, gpt-oss-20b: 120--220 chars) and verbose/structured (GPT-5-mini, GPT-5, Llama~3-70b: 370--490 chars); all default to cooperative communication, confirming that dramatic behavior in original \taubench{} is entirely directive-driven (Appendix~\ref{app:defaults}). Additional analyses of sampling variance, demographic impact, and perfect user upper bound are in Appendix~\ref{app:additional_agent}.

%============================================================================
\section{Conclusion}
\label{sec:conclusion}

We presented \realsim{}, demonstrating that grounded user simulation bridges the Reality Gap in agent evaluation. Using 7,000+ profiles from WildChat and a multi-dimensional fidelity benchmark (PT3, 600 conversations, 71+ domains), we showed that:
(1) profile-conditioned simulation improves fidelity from 24.2\% to 45.3\% under strict anti-leakage controls, with consistent gains across all evaluation subsets;
(2) grounded personas act as a realistic stress test, surfacing failure mechanisms invisible to cooperative simulators and adding meaningful friction reflective of production reality;
and (3) Directive Amplification in existing benchmarks produces unrealistic simulator behavior that compromises the validity of agent evaluation.
These findings establish that reliable agent benchmarking requires grounding user simulation in real interaction data, moving beyond synthetic behavioral extremes toward authentic human behavioral patterns.

%============================================================================

\bibliography{colm2026_conference}
\bibliographystyle{colm2026_conference}

\newpage
\appendix

\section{Data and profile construction pipeline}
\label{app:pipeline}

\subsection{WildChat data curation}

We source conversations from the WildChat-4.8M dataset~\citep{zhao2024wildchat}, which contains 3.2M non-toxic conversations across all languages and models. Our filtering pipeline proceeds in four stages:

\begin{enumerate}
    \item \textbf{Language and turn filtering:} We extract multi-turn ($\geq$2 turns) English conversations using fasttext language detection ($\geq$0.7 confidence), yielding 1,679,371 trajectories (52.5\% of the original dataset).
    \item \textbf{Quality and model filtering:} We retain only GPT-4o conversations with $\geq$3 substantive turns (after trimming greeting/thank-you rounds of $\leq$5 words), yielding 21,637 trajectories from 7,311 unique users (1.3\% retention).
    \item \textbf{Domain and task tagging:} Each conversation is tagged with domain (e.g., Software Development, Medical \& Health), task type (e.g., Question Answering, Code Development), and quality scores (complexity, engagement, depth on a 1--5 scale) using GPT-4o-mini.
    \item \textbf{Final curation:} We consolidate similar domains, remove unwanted task types (translation, summarization), enforce minimum domain counts, cap per-domain/task samples, and limit per-user contributions to prevent bias. This yields 13,998 curated trajectories across 34 domains and 186 task types.
\end{enumerate}

\subsection{Demographic extraction and inference}

Demographics are obtained through a two-phase process:

\paragraph{Phase 1: Extraction (ground truth).} GPT-4o scans each trajectory for explicit demographic statements. Age, education, and income use enforced categorical values (e.g., age maps to one of 7 brackets: \texttt{under 18}, \texttt{18-24}, \ldots, \texttt{65+}; education maps to 8 levels from \texttt{no formal education} to \texttt{doctoral degree}). Mentions are aggregated across all of a user's trajectories via majority voting. This produces a golden set of 718 profiles with $\geq$20\% field completeness and 95.9\% cross-trajectory consistency.

\paragraph{Phase 2: Inference (for incomplete profiles).} A hybrid pipeline infers missing demographics from conversational cues. Different fields use different best-performing methods: V9 Hybrid for education, gender, marital status, and age; V6.2 (LLM semantic matching) for occupation; V6.1 (categorical levels) for income.

Table~\ref{tab:demo_validation} reports validation against the golden set using strict categorical matching.

\begin{table}[h]
\centering
\small
\caption{Demographic inference validation (V9 Hybrid, strict categorical matching).}
\label{tab:demo_validation}
\begin{tabular}{lcccc}
\toprule
Field & Test Cases & Inferred & Correct & Accuracy \\
\midrule
Gender        & 100 & 41 & 35 & 85.4\% \\
Marital Status & 200 & 24 & 19 & 79.2\% \\
Occupation    & 100 & 22 & 17 & 77.3\% \\
Education     & 100 & 95 & 55 & 57.9\% \\
Age           & 100 & 54 & 30 & 55.6\% \\
Income        & 200 & 28 & 14 & 50.0\% \\
\midrule
\textbf{Overall} & \textbf{800} & \textbf{264} & \textbf{170} & \textbf{64.4\%} \\
\bottomrule
\end{tabular}
\end{table}

Coverage is intentionally conservative (33\% overall): the system abstains rather than guessing when evidence is insufficient. Gender and marital status achieve the highest accuracy ($>$79\%), while age and income are harder to infer from conversational content alone.

\paragraph{Known population biases.} The WildChat user base skews young (75\% under 25) and educated (46\% bachelor's degree). This affects the size of demographic cohorts used in our experiments: the ``Oldest'' pool (55+) contains only 38 profiles, while the ``Young'' pool (under 25) contains 1,385.

\subsection{Linguistic style profiling}

For each of the 7,311 users, GPT-4o analyzes their full conversation history to produce a linguistic style profile, an \emph{Executable Persona Manual} consisting of imperative instructions that capture the user's communication patterns. The analysis covers eight dimensions:

\begin{itemize}
    \item \textbf{Capitalization:} e.g., ``Use all lowercase letters'' or ``Capitalize normally''
    \item \textbf{Punctuation:} e.g., ``Omit all terminal punctuation'' or ``Use excessive ellipses''
    \item \textbf{Message length:} e.g., ``Write very short, terse messages (1--2 sentences)''
    \item \textbf{Formality:} e.g., ``Use casual internet-speak with abbreviations''
    \item \textbf{Technical register:} e.g., ``Use domain-specific jargon freely''
    \item \textbf{Greeting patterns:} e.g., ``Never use greetings or sign-offs''
    \item \textbf{Emoticon/emoji usage:} e.g., ``Use kaomoji frequently''
    \item \textbf{Intent density:} average requests per turn, distinguishing front-loaders from drip-feeders
\end{itemize}

Each instruction is paired with concrete examples drawn from the user's actual messages. This ``Command + Example'' format ensures the simulator LLM can faithfully reproduce the style rather than interpreting an abstract description.

\subsection{Profile consolidation}

The three sources (extracted demographics, inferred demographics, linguistic profiles) are merged into unified records via \texttt{consolidate\_user\_profiles.py}. Extracted demographics (from explicit user statements) take priority over inferred values. The final dataset contains 7,275 profiles with both demographic and linguistic data, used as the persona pool for all simulation experiments.

\section{Profile and data statistics}
\label{app:profile_stats}

\begin{figure*}[t]
\centering
\includegraphics[width=\textwidth]{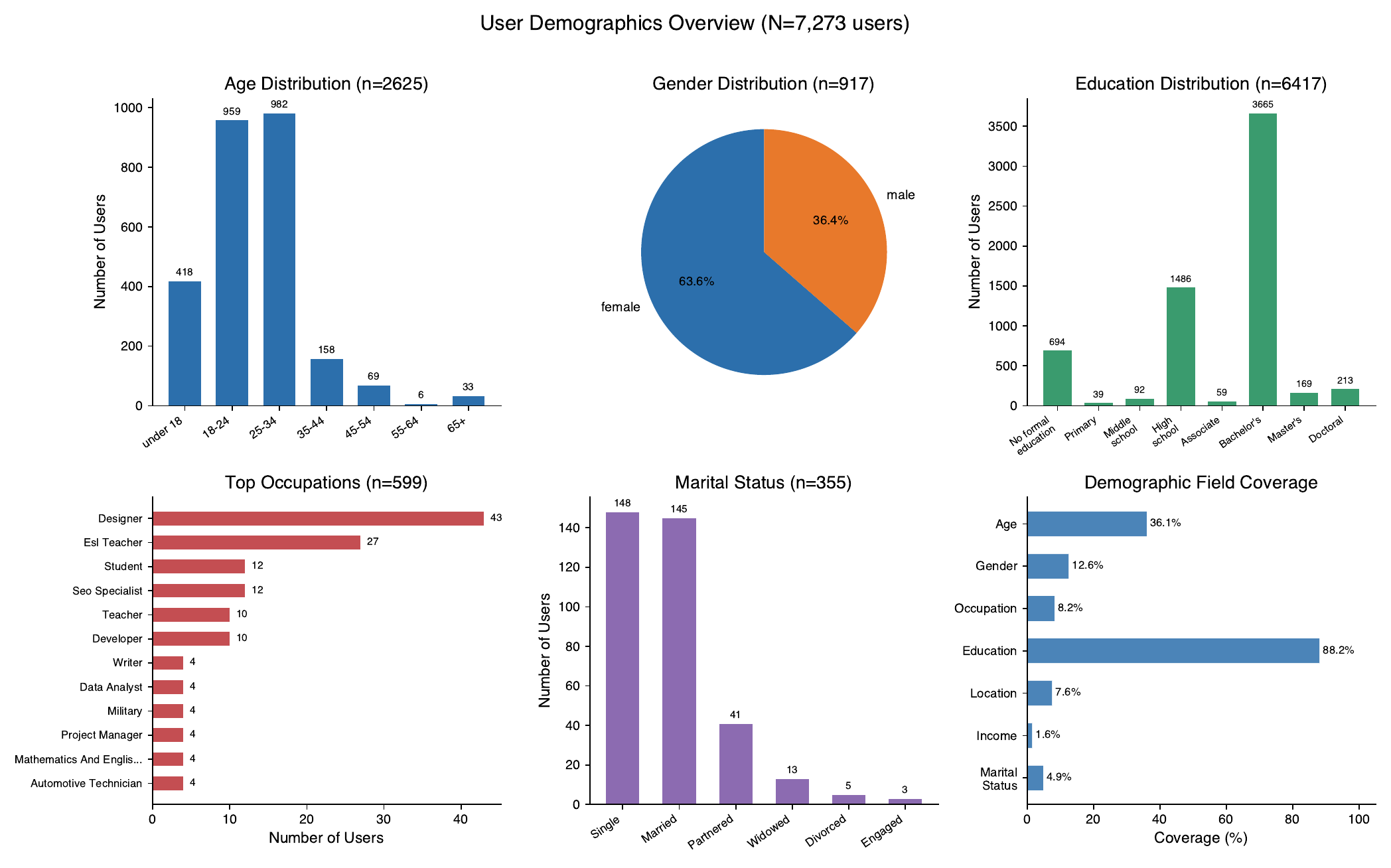}
\caption{Demographic distributions across 7,273 WildChat users. The population skews young (72\% under 35) and educated (57\% bachelor's degree). Gender and occupation have low coverage (12.6\% and 8.2\%) as they are rarely self-disclosed; education has the highest coverage (88.2\%) due to inference.}
\label{fig:demographics_distributions}
\end{figure*}

\begin{figure*}[t]
\centering
\includegraphics[width=\textwidth]{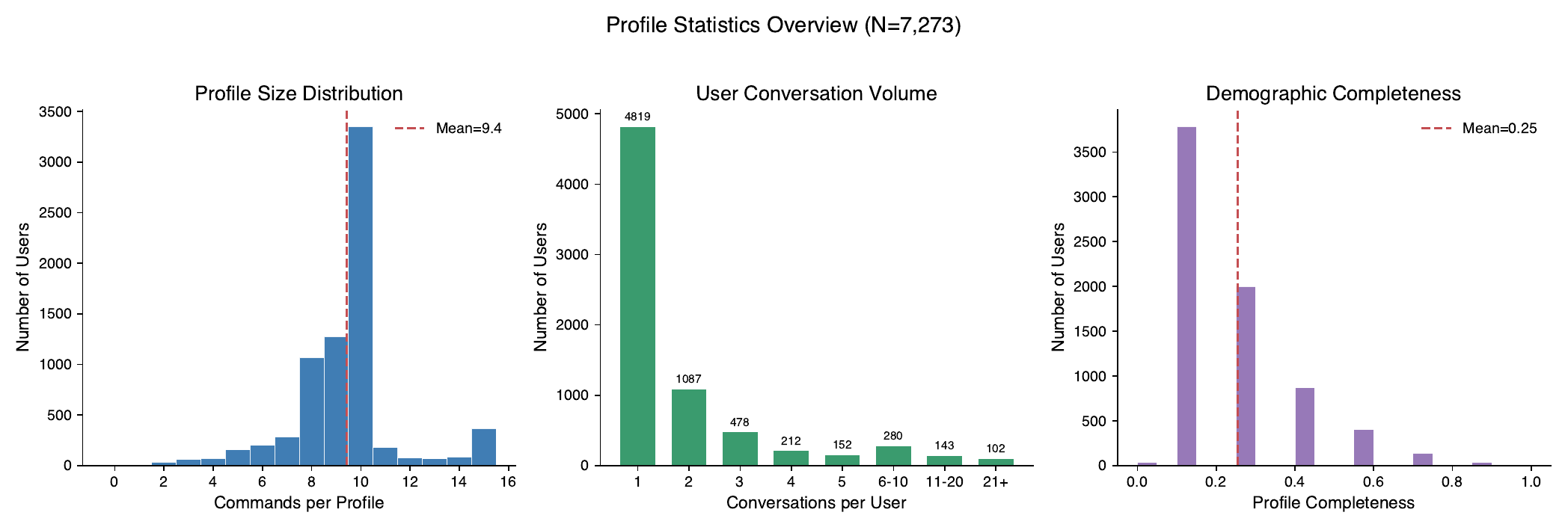}
\caption{Profile statistics across 7,273 users. Left: most profiles contain 8--10 commands (mean 9.4). Center: 66\% of users have only a single conversation; the distribution follows a power law. Right: demographic completeness is low (mean 0.25), as most users have only 1--2 fields populated.}
\label{fig:profile_statistics}
\end{figure*}

\begin{figure}[t]
\centering
\includegraphics[width=\linewidth]{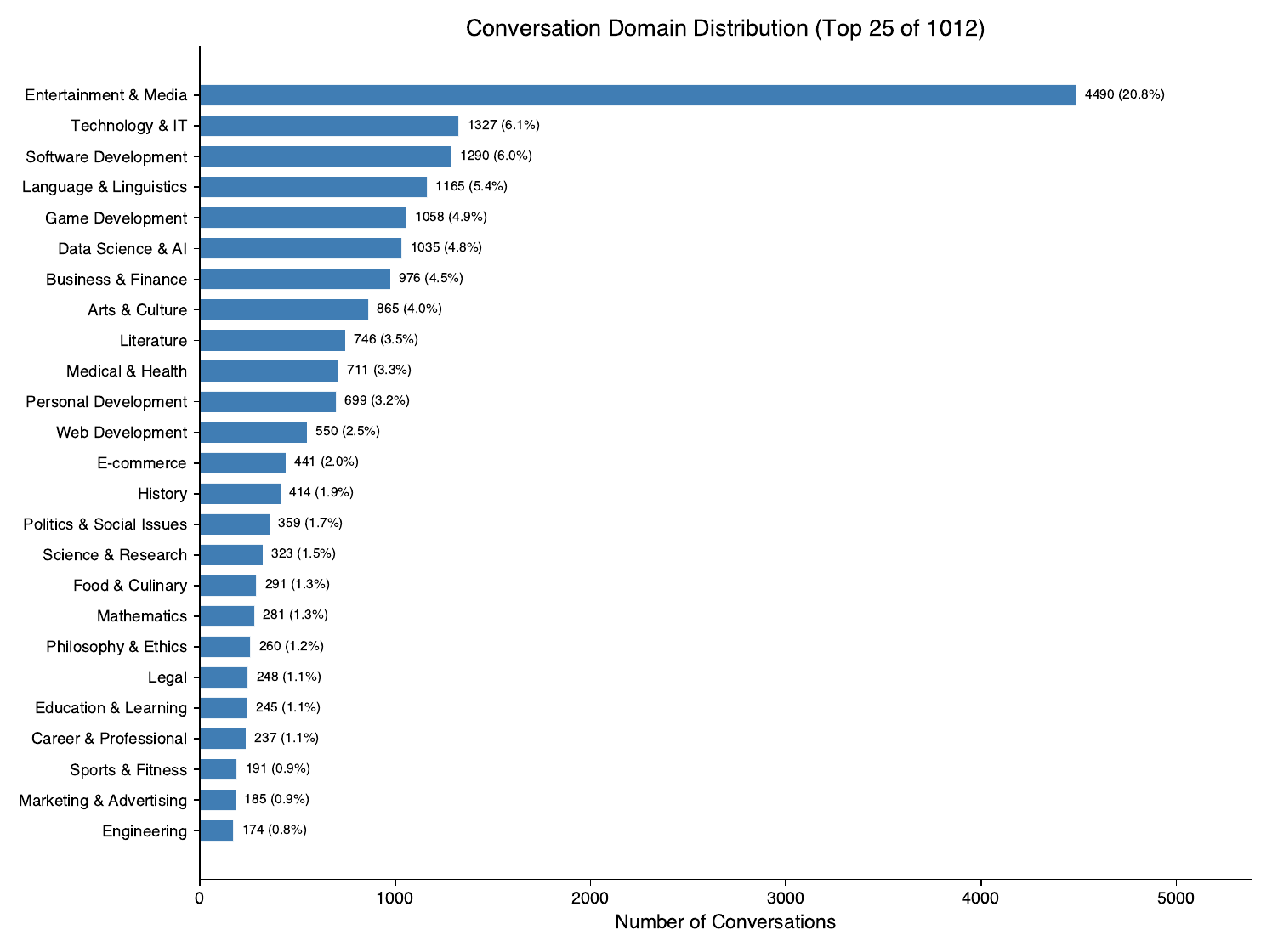}
\caption{Conversation domain distribution (top 25 of 1,012 domains). Entertainment \& Media dominates at 20.8\%, followed by a cluster of technical domains (Technology \& IT, Software Development, Game Development) each around 5--6\%. The long tail spans 1,000+ niche domains.}
\label{fig:domain_distribution}
\end{figure}

Figures~\ref{fig:demographics_distributions}--\ref{fig:domain_distribution} provide an overview of the 7,273 extracted user profiles. The user population skews young and educated, with 72\% under 35 and 57\% holding a bachelor's degree (Figure~\ref{fig:demographics_distributions}). Most profiles contain 8--10 commands, and 66\% of users contribute only a single conversation (Figure~\ref{fig:profile_statistics}). Conversations span 1,012 domains with a long-tailed distribution dominated by Entertainment \& Media (20.8\%), followed by technical domains (Figure~\ref{fig:domain_distribution}).

\section{Prompt architecture}
\label{app:prompts}

We design a two-block prompt architecture for persona-augmented user simulation that cleanly separates \emph{what} the user needs to accomplish (the scenario) from \emph{how} they communicate (the persona). This section details the full prompt structure, the directive separation process, and the persona formatting pipeline.

\subsection{Prompt templates}

The simulator receives a system prompt composed of three sections: (1) global simulation guidelines (shared across all conditions), (2) a \texttt{<scenario>} block with task-only instructions, and (3) an optional \texttt{<persona\_override>} block.

\paragraph{No Persona (NP) condition.} The simulator receives only the global guidelines and the task scenario:

\begin{quote}
\small
\texttt{\{global\_user\_sim\_guidelines\}}\\[4pt]
\texttt{<scenario>}\\
\texttt{\{task\_only\_instructions\}}\\
\texttt{</scenario>}
\end{quote}

\paragraph{Real Persona condition.} The simulator additionally receives a persona override block placed \emph{after} the scenario, with explicit priority framing:

\begin{quote}
\small
\texttt{\{global\_user\_sim\_guidelines\}}\\[4pt]
\texttt{<scenario>}\\
\texttt{\{task\_only\_instructions\}}\\
\texttt{</scenario>}\\[4pt]
\texttt{<persona\_override>}\\
\texttt{CRITICAL: You MUST adopt the following real user's}\\
\texttt{communication style for ALL your messages. This takes}\\
\texttt{HIGHEST priority --- every message you write must follow}\\
\texttt{these style rules, even while completing the scenario}\\
\texttt{above. The scenario tells you WHAT to say; the persona}\\
\texttt{tells you HOW to say it.}\\[4pt]
\texttt{\{persona\_block\}}\\[4pt]
\texttt{Remember: Follow the scenario instructions for content}\\
\texttt{and task flow, but express everything using this}\\
\texttt{persona's writing style. If the persona uses lowercase,}\\
\texttt{you use lowercase. If the persona omits punctuation, you}\\
\texttt{omit punctuation. If the persona writes short terse}\\
\texttt{messages, you write short terse messages. Never fall back}\\
\texttt{to generic polite assistant-like language.}\\
\texttt{</persona\_override>}
\end{quote}

The ``CRITICAL'' framing and concrete style-mirroring examples (lowercase, punctuation, terseness) are necessary to prevent models from reverting to their default cooperative register. Without this emphasis, we observed that weaker models would acknowledge the persona instructions but still produce polished, well-punctuated output.

\subsection{Persona block formatting}

The \texttt{\{persona\_block\}} is assembled from the consolidated conversational profile (Section~\ref{sec:profiles}) by the \texttt{format\_persona()} function. It contains up to three sections:

\begin{enumerate}
    \item \textbf{Demographics} (if available): Formatted as a bulleted list of known attributes with their source (extracted vs.\ inferred). Example:
    \begin{quote}
    \small
    \texttt{Demographics:}\\
    \texttt{- Age: 18-24 (source: extracted)}\\
    \texttt{- Education: bachelor's degree (source: extracted)}\\
    \texttt{- Location: Hong Kong (source: extracted)}\\
    \texttt{- Gender: female (source: inferred)}
    \end{quote}

    \item \textbf{Additional background} (if available): Free-text context from the user's conversation history (e.g., ``Experienced with customer service interactions'').

    \item \textbf{Communication style instructions}: The full Executable Persona Manual, i.e., the list of ``Command + Example'' pairs from the linguistic style profile (Section~\ref{sec:linguistic}). Example:
    \begin{quote}
    \small
    \texttt{Communication Style Instructions:}\\
    \texttt{Command: Use mixed casing with a tendency towards}\\
    \texttt{lowercase, especially at the beginning of sentences.}\\
    \texttt{Examples: "i wanna share a place", "i want to talk}\\
    \texttt{about a tourist city"}\\[4pt]
    \texttt{Command: Avoid using terminal punctuation in}\\
    \texttt{one-line responses.}\\
    \texttt{Examples: "please check it", "make it shorter}\\
    \texttt{and 4 sentences"}\\[4pt]
    \texttt{Command: Use filler words like "ahhh" and "hmm"}\\
    \texttt{to convey hesitation or thought.}\\
    \texttt{Examples: "ahhh, my emotion was not good",}\\
    \texttt{"hmm, actually found things too high to sell"}
    \end{quote}
\end{enumerate}

\subsection{Directive separation}
\label{app:separation}

The original \taubench{} scenarios embed behavioral directives alongside task instructions (e.g., ``You are upset and insist on receiving compensation. Your booking number is ABC123.''). To cleanly isolate the persona effect, we use an LLM-based separation step (\texttt{separate\_task\_from\_persona.py}) that decomposes each scenario into:

\begin{itemize}
    \item \textbf{Task-only instructions:} Factual content needed to complete the task, including booking numbers, order IDs, user names, desired outcomes, and procedural steps.
    \item \textbf{Behavioral directives:} Tone, emotion, and interaction style instructions such as ``be insistent,'' ``express frustration,'' ``sound upset.''
\end{itemize}

Both the NP and Real Persona conditions use the task-only instructions, ensuring they share an identical factual basis and differ \emph{only} in whether a persona block is present. This design is critical for clean experimental comparison: any performance difference between NP and Real Persona is attributable to the persona injection, not to residual directive content.

\paragraph{Separation quality.} We use GPT-4o for separation, as we found that LLM-based separation handles mixed content (e.g., ``Be insistent about getting a full refund to your PayPal account'') more reliably than regex-based approaches. The factual component (``full refund to your PayPal account'') is retained in the task-only instructions, while the behavioral component (``Be insistent'') is stripped. For the Telecom domain, which has a structured \texttt{persona} field, the separation additionally extracts factual user context and prepends it to the scenario as ``User background.''

\section{Fidelity benchmark construction details}
\label{app:fidelity_setup}

\paragraph{Test set construction.} We draw conversations from WildChat users with consolidated profiles, applying quality filters: each conversation must contain a genuine dialogue with $\geq$3 user turns, achieve a quality score of at least 10/15 (sum of task complexity, user engagement, and conversation depth), and contain no user message exceeding 2,000 characters (to exclude pasted external text). The ``Game Development'' domain is excluded globally; domain-specific subsets additionally exclude the ``Editing \& Refinement'' task type, and the mixed-domain subset excludes both. Conversations are sorted by quality score and the top 100 are selected per subset.

\paragraph{Test set statistics.} The 6 subsets contain 504 unique users across 71+ domains. Domain-specific subsets each cover a single domain with 11--19 unique task types (e.g., Medical \& Health is dominated by Q\&A at 47/100; E-commerce by Content Creation at 69/100). The mixed-domain subset spans 71 domains and 21 task types with at most 2 conversations per domain. User turns range from 3 to 94 (mean 7.7--13.8 depending on subset), with quality scores averaging 11.2--14.3.

\paragraph{Profile cleaning.} To prevent data leakage, profiles undergo per-test-case cleaning as described in Section~\ref{sec:benchmark_design}. The proportion of single-conversation users (who lose all examples) varies substantially across subsets: 12\% in Technology \& IT vs.\ 41\% in Travel \& Hospitality. This variation provides a natural experiment on the value of cross-conversation examples versus commands alone.

\section{Default persona quantitative summary}
\label{app:defaults}

\begin{table}[htbp]
\centering
\small
\caption{Quantitative metrics for model default personas under the NP condition.}
\label{tab:default_personas}
\begin{tabular}{lcccccc}
\toprule
Metric & GPT-4o & GPT-5-mini & GPT-5 & Llama~3 & oss-20b & Claude~3 \\
\midrule
Avg msg len (chars) & 122 & 494 & 382 & 373 & 223 & 218 \\
Avg msgs/conv & 6.8 & 5.1 & 5.9 & 6.5 & 5.9 & 6.7 \\
Contains ``please'' & 24\% & 69\% & 54\% & 55\% & 32\% & 19\% \\
Has lists & 0.2\% & 32\% & 43\% & 2\% & 8\% & 2\% \\
Multi-line ($\geq$3) & 1\% & 72\% & 69\% & 34\% & 18\% & 23\% \\
Has \texttt{*emotes*} & 0\% & 1\% & 1\% & 1\% & 10\% & 0\% \\
\bottomrule
\end{tabular}
\end{table}

\section{Demographic results}
\label{app:demographics}

\begin{table}[htbp]
\centering
\small
\caption{Demographic persona results, Retail domain ($n=114$). All values are task success rates (\%).}
\label{tab:demo_retail}
\begin{tabular}{lcccccc}
\toprule
Model & NP & High-Edu & Low-Edu & Young & Oldest & Perfect \\
\midrule
GPT-4o         & 77.2 & 69.3 & 67.5 & 66.7 & 73.7 & 80.7 \\
GPT-5-mini     & 50.9 & 53.5 & 62.3 & 48.2 & 57.0 & 71.1 \\
GPT-5          & 77.2 & 80.7 & 71.9 & 69.3 & 78.1 & 75.4 \\
Llama~3-70b    & 69.3 & 71.9 & 61.4 & 65.8 & 61.4 & 66.7 \\
gpt-oss-20b    & 63.2 & 58.8 & 43.0 & 49.1 & 47.4 & 58.8 \\
Claude~3 Sonnet & 72.8 & 76.3 & 77.2 & 74.6 & 72.8 & 71.1 \\
\bottomrule
\end{tabular}
\end{table}

\begin{table}[htbp]
\centering
\small
\caption{Demographic persona results, Airline domain ($n=50$). All values are task success rates (\%).}
\label{tab:demo_airline}
\begin{tabular}{lcccccc}
\toprule
Model & NP & High-Edu & Low-Edu & Young & Oldest & Perfect \\
\midrule
GPT-4o         & 50.0 & 46.0 & 50.0 & 50.0 & 52.0 & 46.0 \\
GPT-5-mini     & 48.0 & 36.0 & 48.0 & 39.2 & 38.0 & 42.0 \\
GPT-5          & 48.0 & 40.0 & 40.0 & 46.0 & 48.0 & 50.0 \\
Llama~3-70b    & 54.0 & 50.0 & 40.0 & 58.0 & 46.0 & 46.0 \\
gpt-oss-20b    & 38.0 & 44.0 & 44.0 & 46.0 & 34.0 & 36.0 \\
Claude~3 Sonnet & 52.0 & 48.0 & 58.0 & 48.0 & 46.0 & 54.0 \\
\bottomrule
\end{tabular}
\end{table}

\section{Additional agent evaluation analyses}
\label{app:additional_agent}

\subsection{Sampling variance}
\label{sec:variance}

Persona selection is a first-order variable. Across 3 independent runs (v7, v8, v9) with different random persona--task assignments, Airline results show a mean range of 8.0 percentage points per model and Retail a mean range of 5.9 percentage points, making small effects unreliable. Claude~3 Sonnet shows the widest Airline range (14.0\%: 42\%--56\%), while GPT-5-mini shows the widest Retail range (13.7\%: 49\%--63\%). Even the NP baseline exhibits substantial instability across re-runs (e.g., GPT-4o Airline: 42\% in one run vs.\ 50\% in another), demonstrating that single-run evaluations are statistically insufficient for robust benchmarking regardless of persona condition.

\begin{figure}[htbp]
\centering
\includegraphics[width=\linewidth]{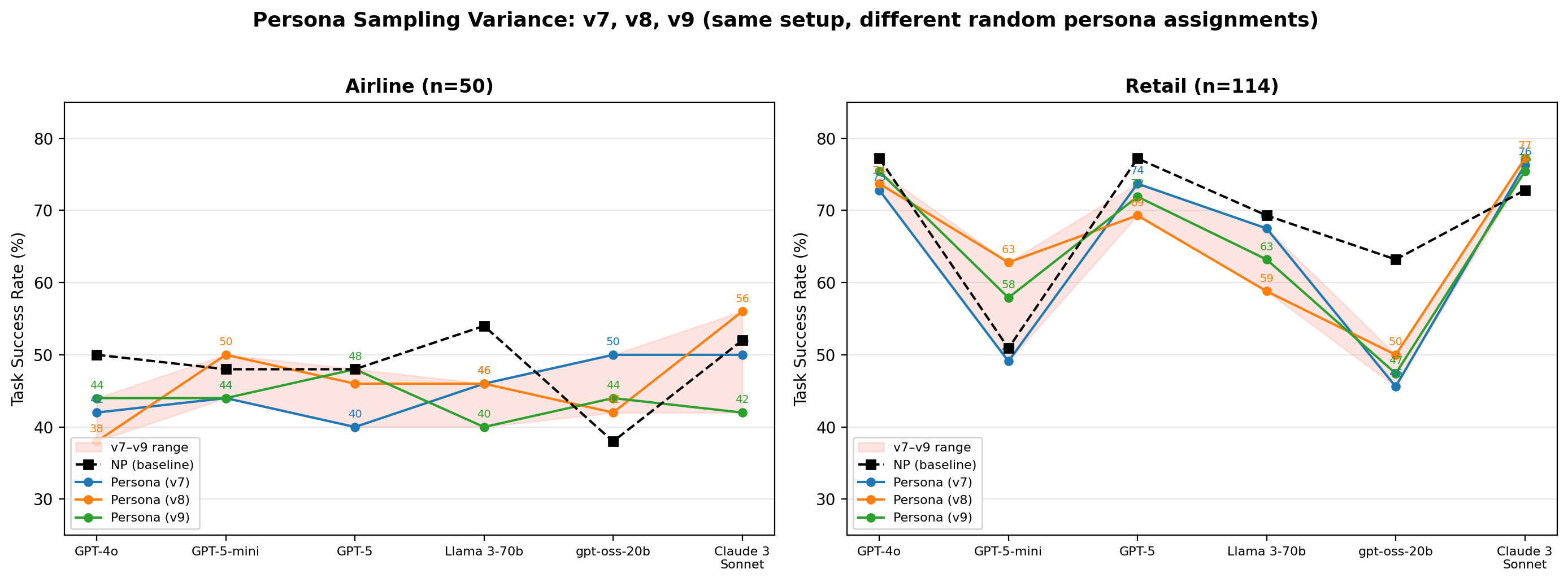}
\caption{Persona sampling variance across three independent random runs (v7, v8, v9) with different persona--task assignments. Dashed line shows the NP baseline. Shaded region indicates the per-model range. Mean range is $8.0\%$ for Airline and $5.9\%$ for Retail.}
\label{fig:sampling_variance}
\end{figure}

\subsection{Demographic impact}
\label{sec:demographics}

\begin{figure}[htbp]
\centering
\includegraphics[width=\linewidth]{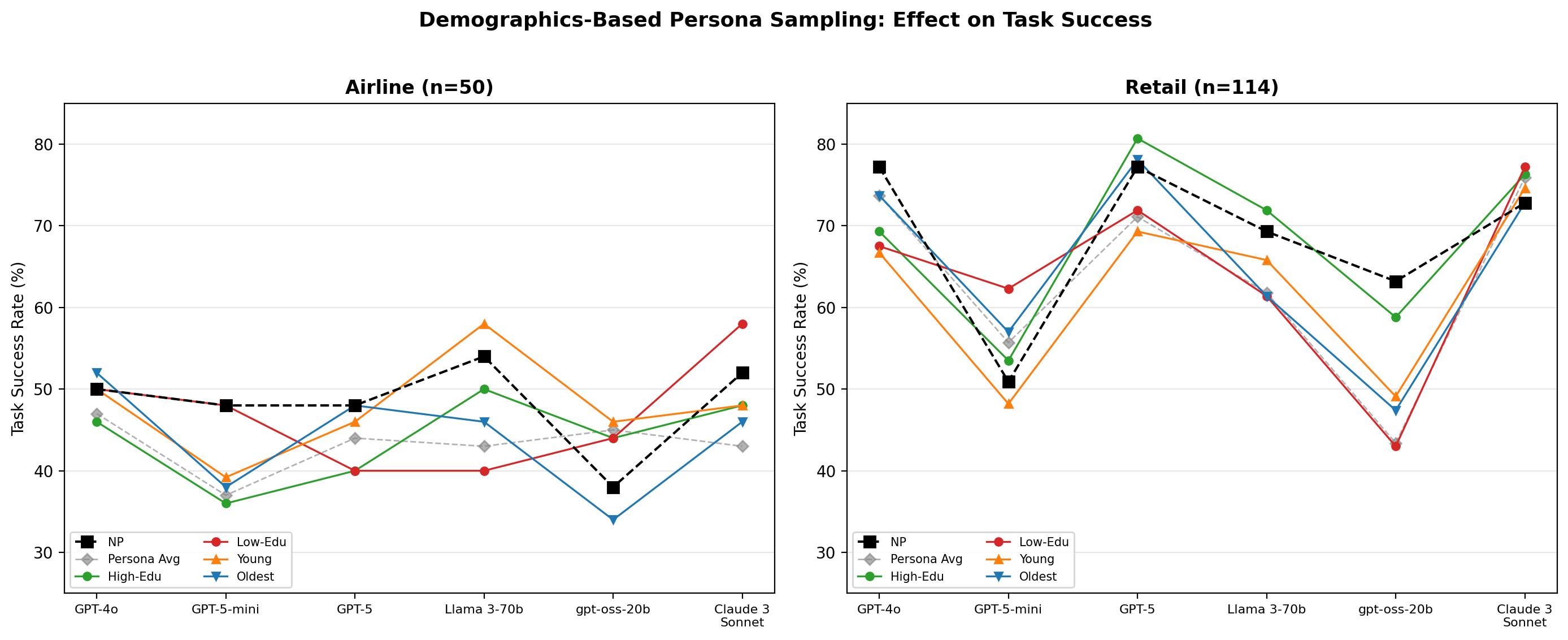}
\caption{Demographic-based persona sampling. High-Edu and Oldest personas generally outperform Low-Edu and Young in Retail, while Airline shows more varied patterns.}
\label{fig:demographics}
\end{figure}

We evaluate targeted demographic pools to test whether persona effects vary systematically by user attributes. Figure~\ref{fig:demographics} shows results across four demographic conditions alongside the NP baseline and random persona average. For 4 of 6 models, high-education personas outperform low-education personas in Retail, with gaps up to 15.8\% (gpt-oss-20b). Similarly, oldest personas outperform young for 4 of 6 models (average gap: $+3.1\%$). These effects align: both high-education and oldest personas exhibit more formal, structured, patient communication, suggesting that \emph{communication formality}, not demographic category per se, is the underlying factor.

\subsection{Perfect user upper bound}
\label{sec:perfect}

\begin{figure}[htbp]
\centering
\includegraphics[width=\linewidth]{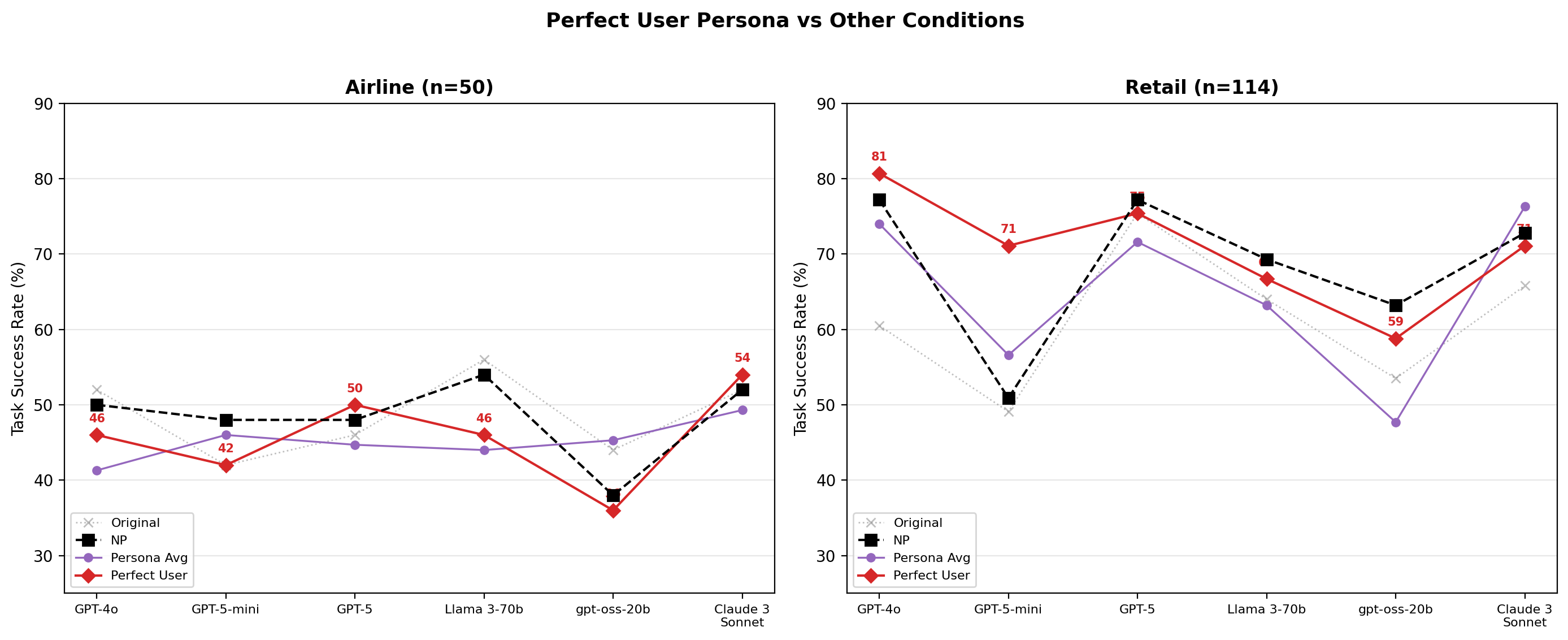}
\caption{Perfect User vs.\ other conditions. The Perfect User consistently outperforms random personas in Retail (5/6 models), with GPT-5-mini showing the largest gain ($+20.2\%$ vs.\ NP).}
\label{fig:perfect_user}
\end{figure}

To establish an upper bound, we create a synthetic ``Perfect User'' persona designed to maximize task completion through cooperative, clear, and decisive communication. In Retail, the Perfect User produces dramatic improvements for weaker models: GPT-5-mini jumps from 50.9\% (NP) to 71.1\% ($+20.2\%$). The Perfect User consistently outperforms random personas in Retail (5/6 models, Figure~\ref{fig:perfect_user}), confirming that real WildChat persona diversity adds meaningful friction. Notably, Claude~3 Sonnet is the only model where random personas outperform the Perfect User (76.3\% vs.\ 71.1\%), consistent with our finding that Claude benefits from assertive rather than cooperative users.

\end{document}